\documentclass[runningheads]{llncs}

\usepackage{graphicx, soul, color, cite, amsmath,amssymb,amsfonts, booktabs}
\usepackage{algorithmic}
\usepackage{graphicx}
\usepackage{textcomp}
\usepackage{xcolor}
\usepackage{subcaption}
\usepackage{lipsum}
\usepackage{listings, tikz, multicol, pgfplots}

\begin{document}

\title{Mixed Reality for Mechanical Design and Assembly Planning}

\author{    Emran Poh  \inst{1} \and 
            Kyrin Liong \inst{2} \and
            Jeannie S.A. Lee\inst{1}
}

\authorrunning{Poh et al.}

\institute{Infocomm Technology, Singapore Institute of Technology, Singapore\\
\and Engineering, Singapore Institute of Technology, Singapore\\
\email{\{emran.poh, kyrin.liong, jeannie.lee\}@singaporetech.edu.sg}}

\maketitle


\begin{abstract}
    
    Design for Manufacturing and Assembly (DFMA) is a crucial design stage within the heavy vehicle manufacturing process that involves optimising the order and feasibility of the parts assembly process to reduce manufacturing complexity and overall cost. Existing work has focused on conducting DFMA within virtual environments to reduce manufacturing costs, but users are less able to relate and compare physical characteristics of a virtual component with real physical objects. Therefore, a Mixed Reality (MR) application is developed for engineers to visualise and manipulate assembly parts virtually, conduct and plan out an assembly within its intended physical environment. Two pilot evaluations were conducted with both engineering professionals and non-engineers to assess effectiveness of the software for assembly planning. Usability results suggest that the application is overall usable (M=56.1, SD=7.89), and participants felt a sense of involvement in the activity (M=13.1, SD=3.3). Engineering professionals see the application as a useful and cost-effective tool for optimising their mechanical assembly designs.
    

\end{abstract}

\section{Introduction}

    The ability to realistically visualise and intuitively manipulate an engineering assembly is key to an efficient design process \cite{jayaram1997virtual}. When presented with a large assembly with heavy components, it is vital to cost-optimise its design and assembly process to mitigate prototyping costs. This `optimisation' is known as DFMA, and improves upon areas such as material selection, machining tolerances, and most importantly the order of assembly of complex components to lower manufacturing and assembly overheads.
    
    Existing work has explored conducting DFMA in virtual environments to reduce physical prototyping, however users are less able to do spatial correlation \cite{jayaram1997virtual, jayaram1999vade, kim2003using}. Spatial correlation refers to the ability to visualise, relate, and compare physical characteristics of a virtual object with real physical objects or environment. Within the context of virtual DFMA, the capability of visualising a virtual assembly within its intended environment is vital for assembly planning. Augmented Reality (AR) has also been utilised to allow virtual assemblies to be overlaid onto physical environments for visualisation, although with limited interaction \cite{poupyrev2001tiles, zauner2003authoring}. Mixed Reality (MR) presents a useful platform where engineers will be able to intuitively interact, conduct design reviews without prototyping, and visualise their virtual Computer-Aided Design (CAD) assemblies within their intended physical environment.
    
    Understanding the requirements of conducting an assembly within a mixed reality environment through a requirements gathering session, the goal is to develop an intuitive and effective MR application for conducting assembly planning. Two pilot studies were conducted to quantitatively and qualitatively evaluate the usability of the proposed application and gather useful insight into its effectiveness and learnability.

\section{Related Work}

    Existing virtual assembly planning systems allow users to be able to visualise and evaluate the assembly of their mechanical systems in various virtual environments with different tools. External controllers such as a mouse \cite{ganier2014evaluation} or a haptic device \cite{gallegos2017analysis} provide a physical object for the user to interact with the virtual environment, however will incur cost for setting up the tracking system for each device. VR-based systems \cite{wolfartsberger2019analyzing, seth2011virtual,jayaram1997virtual, chen2020} provide a useful platform for simulating different environments, however it is limited in its ability to replicate dynamic and realistic environments. AR systems \cite{poupyrev2001tiles} addresses the lack of spatial correlation through overlaying useful information onto physical surfaces, however the virtual objects do not interact with the physical world. While conducting engineering design reviews in a virtual environment shortens the product development life-cycle and lowers manufacturing overheads \cite{jayaram1999vade, poh2021}, these systems lack the spatial awareness needed to conduct feasibility tests of comparing the virtual models within a real physical environment. Mixed reality presents a platform in which virtual objects could interact with the physical world through spatial mapping and gesture interaction. Through conducting assembly planning within MR, the user will be able to visualise and iterate on their designs quickly, intuitively interact with their virtual models, and reduce manufacturing and prototyping overheads.

\section{Assembly Planning Features}

    The goal of this work is to utilise MR as a platform to conduct engineering design reviews. A list of assembly planning features were obtained through a requirement gathering session with professional design engineers. Within the list, three primary categories of features were identified: \textit{Component Manipulation}, \textit{Assembly Guidance}, \textit{Spatial Awareness}.

    \begin{table}[b!]
          \caption{List of Assembly Planning Features}
          \label{tab:fss}
          \centering
            \begin{tabular}{l l l}
            \toprule
            Category & \hphantom{A} & Feature \\ \midrule
            Component & & Assembly Component Spawning \\
            Manipulation & & Assembly Component Browsing \\ 
             & & Manipulation (Single-Axis Rotation, Translation) \\ \midrule
            Spatial & & Spatial Anchoring \\
            Awareness & & Orientation Indicator \\
             & & Directional Indicator \\ \midrule
            Assembly & & Component Positioning Guidance \\
            Guidance & & Audio Feedback \\
             & & Component-to-Component Collision \\ 
            \bottomrule
            \end{tabular}
    \end{table}


    \begin{figure}[t!]
        \centering
        \begin{subfigure}[t]{.58\textwidth}
          \centering
          \includegraphics[height=3.3cm]{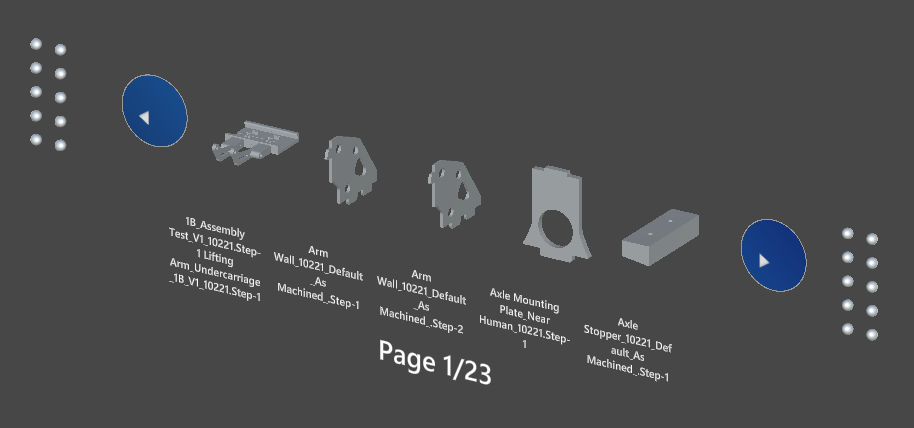}
          \caption{\label{fig:carousel}}
        \end{subfigure}%
        \hfill
        \begin{subfigure}[t]{.42\textwidth}
          \centering
          \includegraphics[height=3.3cm]{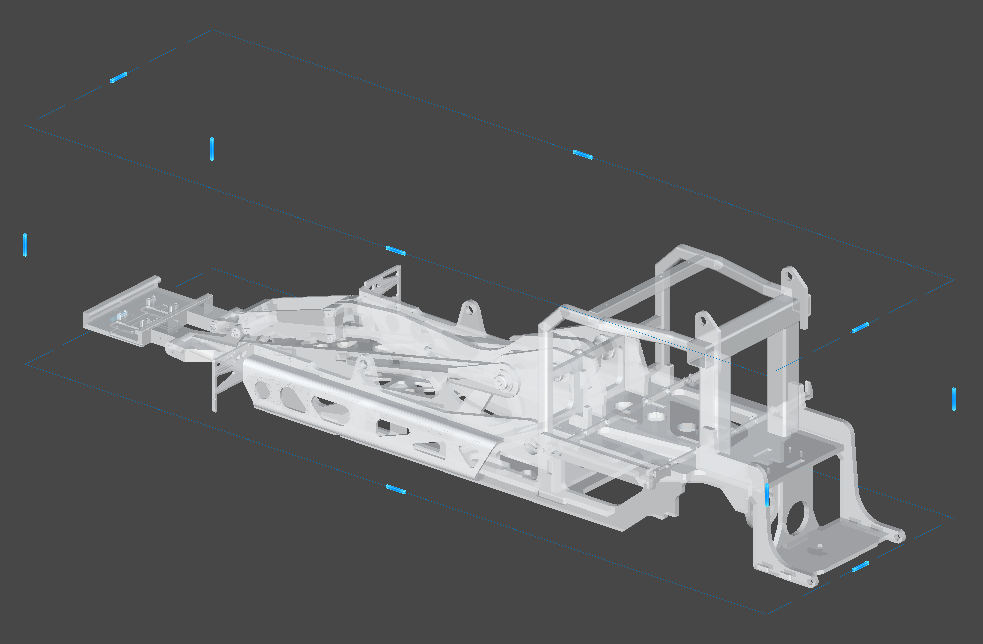}
          \caption{\label{fig:bounds}}
        \end{subfigure}
        \caption{Assembly Planning Features. \subref{fig:carousel} Component Browsing interface, \label{fig:bounds}`Bounds' interface for Anchoring}
    \end{figure}
    
    \subsubsection{Component Manipulation}
    Being able to manipulate assembly component parts is critical in testing and planning out possible manoeuvres for assembly. A 3D browsing interface (Fig \ref{fig:carousel}) enables the user to spawn and hide component objects to allow the user to bring parts into the assembly area when the user requires of it. A toggle button from the 'Hand Menu', the user will be able to toggle between freehand bi-manual manipulation and single-axis rotation. 
    
    \subsubsection{Spatial Awareness} 
     Spatial awareness is critical in developing a useful MR application, especially within the context of an engineering design review. Spatial anchoring was added such that the virtual assembly model would be able to adhere to a specific location in the real physical environment to improve transfer of knowledge and correlation. This anchoring also helps in visualising the 1:1 scale of the model in real life.  To provide a frame of reference for assembly orientation, the Orientation Indicator rotates along with the user in relation to the anchored assembly. Upon spawning a component, an arrow will appear around the component pointing towards the direction of its position within the anchored assembly.
     
    \subsubsection{Assembly Guidance}
    
    To provide assembly guidance, a list of visual and audio prompts were introduced to assist during assembly planning. Upon manipulation of an assembly component, a translucent copy of the same component appears in the position of where the component should be within the anchored assembly. Component-To-Component collision is an effective tool to definitively validate assembly manoeuvres. To simulate this interaction, the components change colour from a base colour of white (indicating no collision) to blue (indicating collision). Audio feedback emphasises any application events such as button presses, component collisions, and manipulation.

\section{Implementation}

    \begin{figure}[h]
        \centering
        \begin{subfigure}[t]{.41\textwidth}
          \centering
          \includegraphics[height=2.7cm]{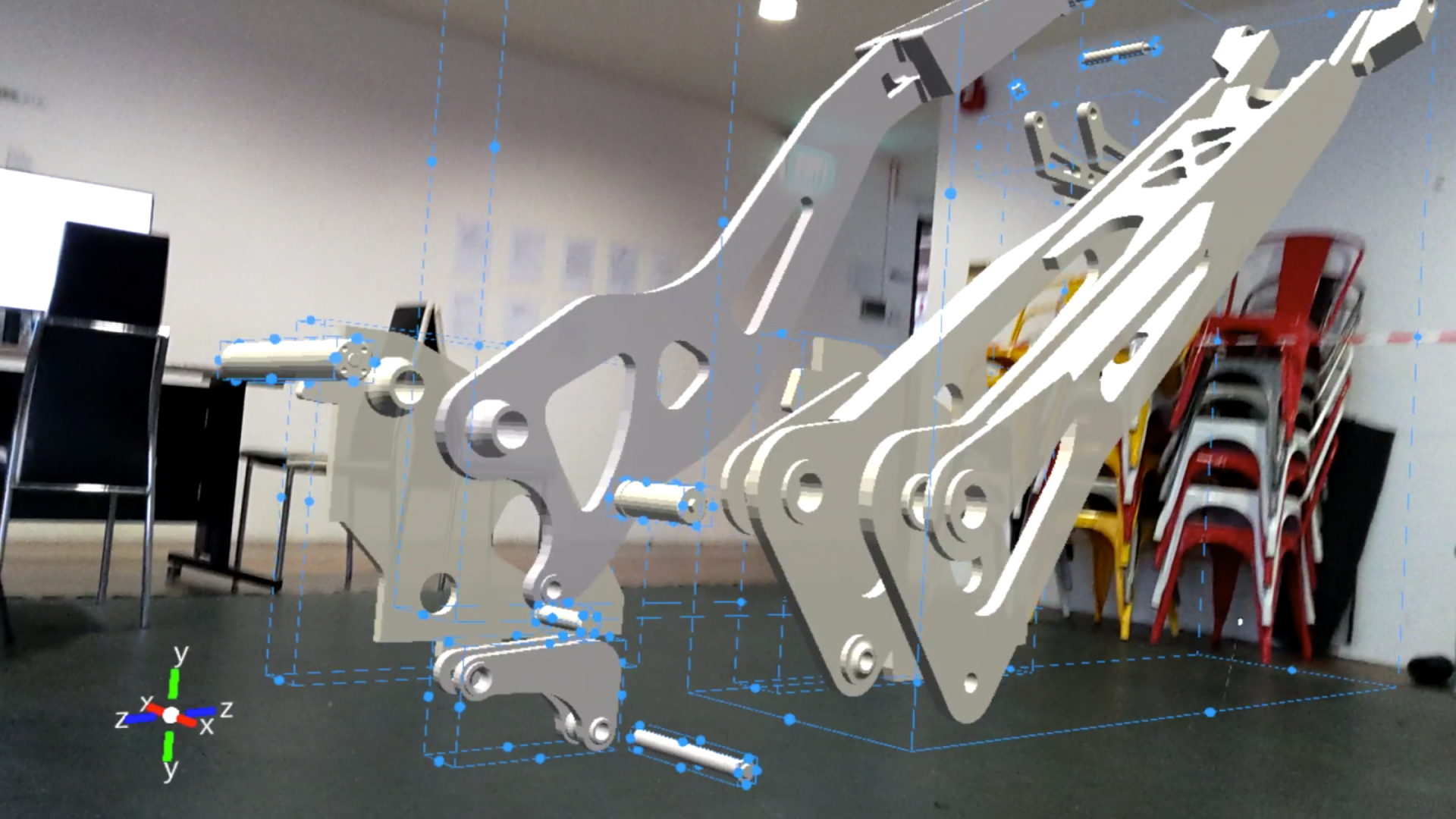}
          \caption{\label{fig:ol_mr}}
        \end{subfigure}%
        \hfill
        \begin{subfigure}[t]{.215\textwidth}
          \centering
          \includegraphics[height=2.7cm]{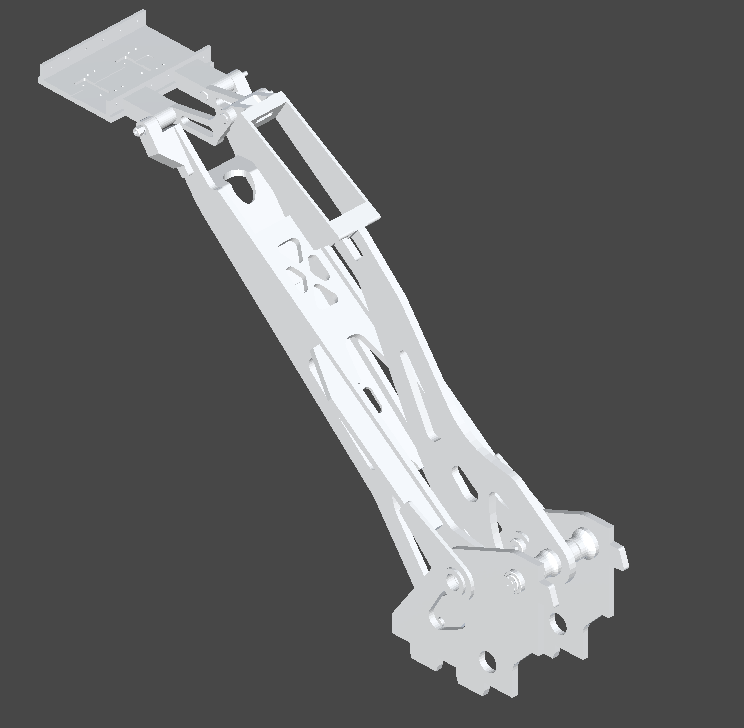}
          \caption{\label{fig:ol_assembled}}
        \end{subfigure}%
        \hfill
        \begin{subfigure}[t]{.37\textwidth}
          \centering
          \includegraphics[height=2.7cm]{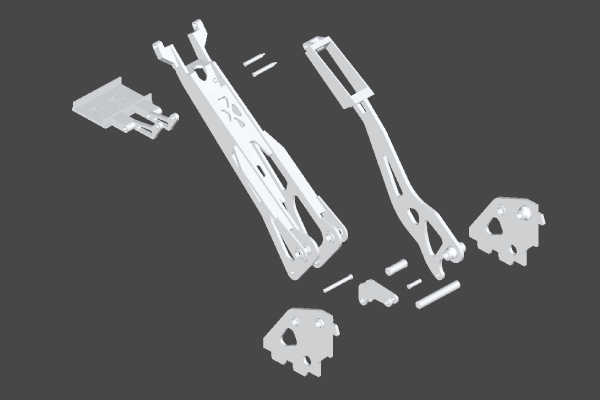}
          \caption{\label{fig:ol_exploded}}
        \end{subfigure}
        \caption{Vehicular Heavy Lifter with 13 parts (13 parts). (\subref{fig:ol_mr}) View of Assembly Components with Orientation Indicator within MR View, (\subref{fig:ol_assembled}) shows a completed view of the assembly, (\subref{fig:ol_exploded}) shows an exploded view of the assembly}
        \label{fig:ol}
    \end{figure}
    
    All assembly planning features and interactions were built using gestures and interaction components within Microsoft's Mixed Reality Toolkit (MRTK). The application was developed using Unity Engine and deployed and tested on Microsoft Hololens 2. For the CAD models, the assembly model used were designed and tested through SolidWorks and converted into a Unity-compatible format using Blender. The subject (Figure \ref{fig:ol}) is a large assembly of a vehicular heavy lifter (Dimensions: 3.5m x 1.5m x 2m, Actual Physical Assembly Weight: 4 tons).
    
    
    \begin{figure}
        \centering
        \begin{subfigure}{.49\textwidth}
          \centering
            \includegraphics[height=2.9cm]{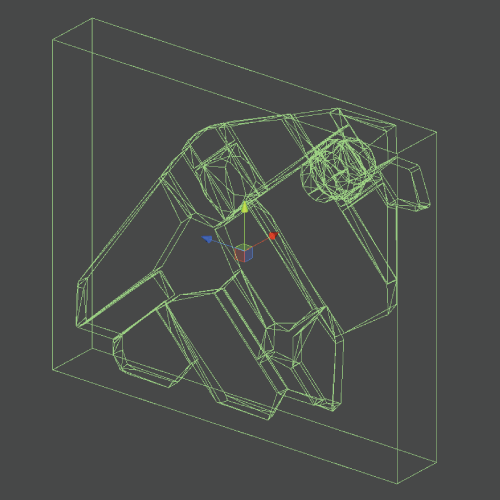}
            \includegraphics[height=2.9cm]{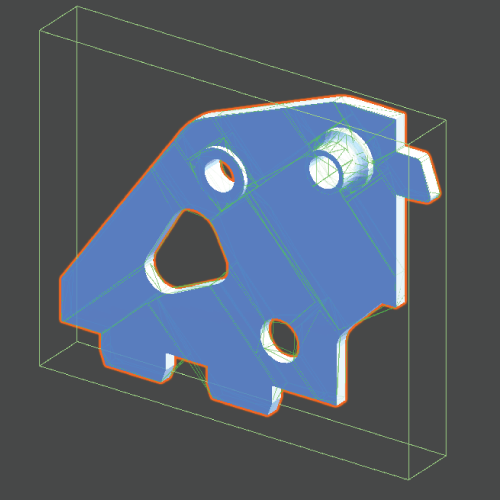}
            \caption{\label{fig:colliders}}
        \end{subfigure}%
        \hfill
        \begin{subfigure}{.49\textwidth}
          \centering
            \includegraphics[height=2.9cm]{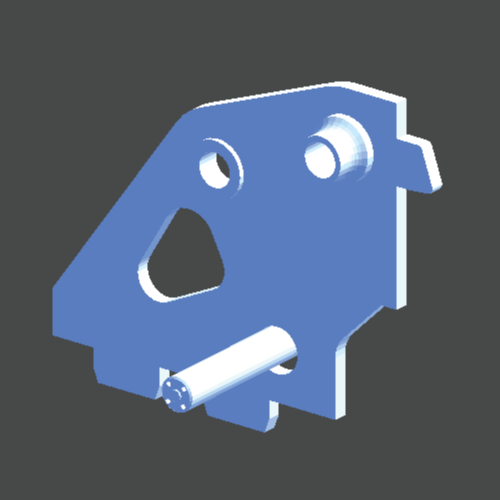}
            \includegraphics[height=2.9cm]{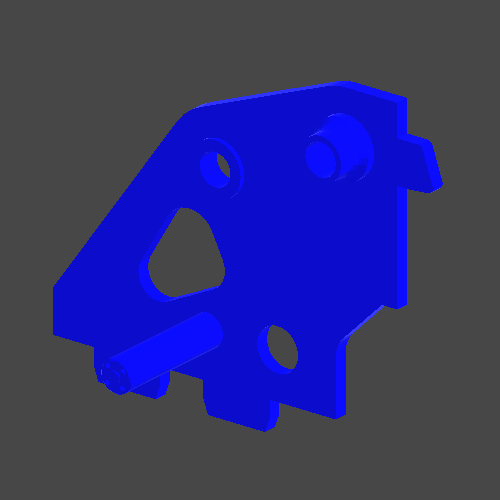}
            \caption{\label{fig:colliders}}
        \end{subfigure}
        \caption{Component Collision Collider Implementation. (\subref{fig:colliders}) (From Left to Right) Box Collider and the generated Non-Convex Collider. Base Component with Box Collider. (\subref{fig:colliders}) (From Left to Right) Two solid assembly components in varying levels of collision (No Collision, Colliding)}
        \label{fig:collision}
    \end{figure}
    
    The user is able to carry out several basic interactions for manipulation and rotation through the attachment of \textit{ManipulationHandler} script for input detection and handling. For single axis rotation, the \textit{Bounds} component provides a visible interface which users could interact with to pivot the model along its axes. To simulate collision, a non-convex collider is generated and attached to each assembly component upon spawning. Upon contact between components, a collision event will be triggered which will modify the colour of the material of each colliding components. To avoid false collisions caused by overlapping and jagged colliders, a user-defined parameter was introduced to define a margin of error that the colliders could overlap before triggering a collision event.

\section{Evaluation}

    
    To evaluate the effectiveness of the application as a platform for assembly planning, two preliminary studies were conducted. First, the relative usability and sense of flow of the application was assessed by a group of users with little to no existing expertise through a pilot quantitative study. Second, a pilot qualitative study was conducted on a group of senior engineers to evaluate the effectiveness of the application for conducting a an assembly planning.
    
    \subsection{Pilot Quantitative Study and Results}
    
    \begin{table}[b!]
        \caption{System Usability Scale Scores (n=7)}
          \label{tab:sus}
        \centering
        \begin{tabular}{llcc}
            \toprule
            No. & Question & Mean & SD \\ \midrule
            1 & I think I would like to use this interface frequently & 4.29 & 0.76 \vspace{.8ex} \\ 
            2 & I found the interface unnecessarily complex & 3.86 & 0.38 \vspace{.8ex} \\ 
            3 & I thought the interface was easy to use & 4.29 & 0.49 \vspace{.8ex} \\ 
            4 & \begin{tabular}[c]{@{}l@{}}I think that I would need the support of a technical person to be\\ able to use this interface\end{tabular} & 2.86 & 1.21 \vspace{.8ex} \\ 
            5 & I found the various functions in this interface were well-integrated & 4.57 & 0.53 \vspace{.8ex} \\ 
            6 & I thought there was too much inconsistency in this interface & 3.86 & 0.90 \vspace{.8ex} \\ 
            7 & \begin{tabular}[c]{@{}l@{}}I would imagine that most people would learn to use this interface\\ very quickly\end{tabular}  & 3.71 & 0.49 \vspace{.8ex} \\ 
            8 & I found the interface very cumbersome to use. & 4.29 & 0.49 \vspace{.8ex} \\ 
            9 & I felt very confident using the interface & 4.00 & 0.58 \vspace{.8ex} \\ 
            10 & \begin{tabular}[c]{@{}l@{}}I needed to learn a lot of things before I could get going with\\ this interface\end{tabular}  & 3.57 & 1.40 \\ \midrule
            \multicolumn{2}{l}{Usability Score} & 56.07 & 7.89 \\
            \bottomrule
        \end{tabular}
    \end{table}
    
    In this study, participants (n=7, Age: 20-45) who have no prior experience in mixed reality and engineering are required to assemble the vehicular lifter through the use of all available assembly guidance. At the end of the assembly session, the relative usability \cite{brooke1986system} and perceived flow \cite{csikszentmihalyi1990flow} of the application were recorded.
    
    The usability results from the experiment in Table \ref{tab:sus} suggest that the interface is empirically usable (M=56.07, SD=7.89). Given the lack of prior experience in engineering or mixed reality, this could suggest that more visual instruction may be necessary for first-time users to navigate the application. Findings also show in Table \ref{tab:fss} that the participants felt that they were clear about their goals (M=12.29, SD=1.60), had a sense of control when experiencing the application (M=11.57, SD=1.51), and their thoughts ran somewhat smoothly with minor difficulty in concentrating (M=17.00, SD=2.16). However, the participants felt that there was too much challenge in understanding the interface and that they feel that they were somewhat competent (M=13.1, SD=3.3). Despite this, most participants reported they enjoyed the experience (M=6.57, SD=1.13).
    
    \begin{table}[ht!]
      \caption{Flow State Scale Scores (n=7)}
          \label{tab:fss}
          \centering
            \begin{tabular}{llcc}
            \toprule
            No. & Question & Mean & SD \\ \midrule
            1 & I was not confused about my objectives & 6.29 & 0.76 \\
            2 & I knew what I had to do each step of the way & 6.00 & 1.15 \\
            3 & My thoughts/activities ran fluidly and smoothly & 5.14 & 1.57 \\
            4 & I had no difficulty concentrating & 5.43 & 1.40 \\
            5 & I was totally absorbed in what I was doing & 6.43 & 0.53 \\
            6 & I feel just the right amount of challenge & 3.57 & 1.51 \\
            7 & I felt that I must not make any mistakes & 4.43 & 1.72 \\
            8 & I think that my competence in this area was & 5.14 & 1.21 \\
            9 & The right thoughts/movements occurred of their own accord & 5.71 & 0.49 \\
            10 & I felt that I had everything under control & 5.86 & 1.21 \\
            11 & I enjoyed the experience & 6.57 & 1.13 \\ \midrule
            \multicolumn{2}{l}{Subscale 1: Clear Goals (Qns 1, 2)} & 12.29 & 1.60 \\
            \multicolumn{2}{l}{Subscale 2: Action-Awareness Merging (Qns 3, 4, 5)} & 17.00 & 2.16 \\
            \multicolumn{2}{l}{Subscale 3: Challenge Skill Balance (Qns 6, 7, 8)} & 13.14 & 3.34 \\
            \multicolumn{2}{l}{Subscale 4: Sense of Control (Qns 9, 10)} & 11.57 & 1.51 \\
            \multicolumn{2}{l}{Subscale 5: Autotelic Experience (Qn 11)} & 6.57 & 1.13 \\
            \bottomrule
        \end{tabular}
    \end{table}

    \subsection{Pilot Qualitative Study and Results}
    
    A pilot think-aloud was conducted on senior design engineers (n=3, Age: 35-50) to assess the feasibility of the features developed in the application. The participants had 8 to 10 years of experience in coordinating assembly planning and conducting design reviews. The participants were required to assemble the heavy vehicular loader. A thematic analysis was conducted upon the think-aloud and semi-structured feedback by the participants.


    During the walkthrough, participants mentioned that finger-pointing and pinch-tap as a tool for basic manipulation `feels' natural however compromises on accuracy that is needed for more complex manoeuvres. While there were visible assembly guidance prompts, participants suggested that more guidance is needed when conducting an assembly in a mixed environment. While collision exists in real life, other physical information such as assembly tolerance and sub-assembly information could be useful. Overall, participants mentioned the suitability of the application for macro-structure assembly planning and highlighted its potential for visualising and validating possible sub-assembly interactions involved without the need for physical prototyping.
    
    \subsection{Evaluation Summary}
    
     Quantitative results show that while the interface can be initially challenging, users were eventually able to find enjoyment and a sense of control when navigating. For engineering professionals, this could hint at a steeper learning curve when first introduced to concepts that require specialised knowledge in an unfamiliar environment. Qualitative results suggest that engineering professionals find part collisions and superposition of virtual assembly parts within a physical context as a useful and cost-effective tool for conducting an assembly design review. A further study could be conducted on a larger population of engineering experts which would yield greater accuracy in understanding the effectiveness of the application.
    
\section{Conclusion}

    The developed application is designed to enhance the efficiency of mechanical design engineers when conducting DFMA. MR presents a compelling platform for testing and visualising complex virtual assembly models and component interactions onto physical space without the need for prototyping. Results from the preliminary studies reveal that the application is usable and accessible despite a learning curve. Feedback from design engineers highlights that they foresee a potential for using MR in improving their DFMA process. Future work could involve exploring a multi-user experience for assembly planning to correlate better with the nature of assembly planning in a group.
    
\section*{Acknowledgements}
The authors of this paper wish to acknowledge the contributions provided by the technical staff of HelloHolo and Hope Technik Pte Ltd Singapore. This research was supported by the Singapore Institute of Technology Ignition Grant (R-MOE-E103-G001).

\bibliographystyle{splncs04}
\bibliography{paper}

\begin{thebibliography}{10}
\providecommand{\url}[1]{\texttt{#1}}
\providecommand{\urlprefix}{URL }
\providecommand{\doi}[1]{https://doi.org/#1}

\bibitem{brooke1986system}
Brooke, J.: System usability scale (sus): A quick-and-dirty method of system
  evaluation user information. digital equipment co ltd. Reading, UK pp.~1--7
  (1986)

\bibitem{chen2020}
Chen, Q., Low, S.E., Yap, J.W., Sim, A.K., Tan, Y.Y., Kwok, B.W., Lee, J.S.,
  Tan, C.T., Loh, W.P., Loo, B.L., Wong, A.C.: Immersive virtual reality
  training of bioreactor operations. In: 2020 IEEE International Conference on
  Teaching, Assessment, and Learning for Engineering (TALE). pp. 873--878
  (2020). \doi{10.1109/TALE48869.2020.9368468}

\bibitem{csikszentmihalyi1990flow}
Csikszentmihalyi, M., Abuhamdeh, S., Nakamura, J., et~al.: Flow (1990)

\bibitem{gallegos2017analysis}
Gallegos-Nieto, E., Medell{\'\i}n-Castillo, H.I., Gonz{\'a}lez-Badillo, G.,
  Lim, T., Ritchie, J.: The analysis and evaluation of the influence of
  haptic-enabled virtual assembly training on real assembly performance. The
  International Journal of Advanced Manufacturing Technology  \textbf{89}(1),
  581--598 (2017)

\bibitem{ganier2014evaluation}
Ganier, F., Hoareau, C., Tisseau, J.: Evaluation of procedural learning
  transfer from a virtual environment to a real situation: a case study on tank
  maintenance training. Ergonomics  \textbf{57}(6),  828--843 (2014)

\bibitem{jayaram1997virtual}
Jayaram, S., Connacher, H.I., Lyons, K.W.: Virtual assembly using virtual
  reality techniques. Computer-aided design  \textbf{29}(8),  575--584 (1997)

\bibitem{jayaram1999vade}
Jayaram, S., Jayaram, U., Wang, Y., Tirumali, H., Lyons, K., Hart, P.: Vade: a
  virtual assembly design environment. IEEE computer graphics and applications
  \textbf{19}(6),  44--50 (1999)

\bibitem{kim2003using}
Kim, C.E., Vance, J.M.: Using vps (voxmap pointshell) as the basis for
  interaction in a virtual assembly environment. In: International Design
  Engineering Technical Conferences and Computers and Information in
  Engineering Conference. vol. 36991, pp. 1153--1161 (2003)

\bibitem{poh2021}
Poh, E., Liong, K., Lee, J.S.A.: Mixed reality interface for load application
  in finite element analysis. In: Meiselwitz, G. (ed.) Social Computing and
  Social Media: Experience Design and Social Network Analysis. pp. 470--483.
  Springer International Publishing, Cham (2021)

\bibitem{poupyrev2001tiles}
Poupyrev, I., Tan, D.S., Billinghurst, M., Kato, H., Regenbrecht, H.,
  Tetsutani, N.: Tiles: A mixed reality authoring interface. In: Interact.
  vol.~1, pp. 334--341. Citeseer (2001)

\bibitem{seth2011virtual}
Seth, A., Vance, J.M., Oliver, J.H.: Virtual reality for assembly methods
  prototyping: a review. Virtual reality  \textbf{15}(1),  5--20 (2011)

\bibitem{wolfartsberger2019analyzing}
Wolfartsberger, J.: Analyzing the potential of virtual reality for engineering
  design review. Automation in Construction  \textbf{104},  27--37 (2019)

\bibitem{zauner2003authoring}
Zauner, J., Haller, M., Brandl, A., Hartman, W.: Authoring of a mixed reality
  assembly instructor for hierarchical structures. In: The Second IEEE and ACM
  International Symposium on Mixed and Augmented Reality, 2003. Proceedings.
  pp. 237--246. IEEE (2003)

\end{thebibliography}

\end{document}